\documentclass[10pt]{iopart}

\usepackage{graphicx}
\usepackage{supertabular}
\usepackage{iopams}

\newcommand{\eqref}[1]{(\ref{#1})}
\newcommand{\text}[1]{{\rm #1}}

\begin{document}

\title{Comment on `The Newtonian force experienced by a point mass near a finite cylindrical source'}

\author{M Azreg-A\"{\i}nou}

\address{Ba\c{s}kent University, Department of Mathematics, Ba\u{g}l\i ca Campus, Ankara, Turkey}
\begin{abstract}
We compare the computation times and precisions of three expansions for the gravitational potential. The evaluation of the series by Selvaggi \etal (2008 \textit{Class. Quantum Grav.} \textbf{25} 015013) is time-consuming to be applied to the gravitational constant ($G$) or STEP experiments.\\

\end{abstract}


We compare the computational efficiency of three published expansions for the gravitational potential (GP) of a finite, hollow, thick-walled cylinder with constant mass density $\rho$ and reflection symmetry with respect to the plane $z=0$. The expansions are those developed by Lockerbie \etal, Schlamminger \etal and Selvaggi \etal~\cite{L,PRD,S}.

If $V$ denotes the GP/($2\pi G\rho$) and $(2L,a,b)$ the length, inner and outer radii of the cylinder, Lockerbie \etal formula, which is valid outside the sphere  (S) of radius $\sqrt{b^2+L^2}$ and center at the origin, writes in terms of Legendre polynomials $P_{2n}$ as
\begin{equation}\label{L1}
\hspace{-21mm} V_{\text{Loc}}(r,z) = \sum_{n=0}^\infty \left\{\sum_{p=0}^n \frac{(-1)^p(2n)!L^{2n-2p+1}(b^{2p+2}-a^{2p+2})}{2^{2p}p!(p+1)!(2n-2p+1)!}\right\}
    \frac{P_{2n}(z/\sqrt{r^2+z^2})}{(r^2+z^2)^{n+(1/2)}} \,,
\end{equation}

The GP, $V_{\text{Sch}}(r,z)$, by Schlamminger \etal formula is expressed in terms of the GP along the axis of symmetry $V(0,z)$. If we introduce $f(x)=x+\sqrt{b^2+x^2}$ and $g(x)=x\sqrt{b^2+x^2}$, then $V(0,z)$ \& $V_{\text{Sch}}(r,z)$ are evaluated by
\begin{equation}\label{A11}
\hspace{-21mm} 2V(0,z)=b^2\{\ln[f(z+L)]-\ln[f(z-L)]\} + g(z+L) - g(z-L)
    - (a\rightarrow b)\,,
\end{equation}
\begin{equation}\label{Af}
\hspace{-21mm} V_{\text{Sch}}(r,z)=V(0,z)+\sum_{n=1}^\infty \frac{(-1)^nV^{(2n)}(0,z)}{2^{2n}(n!)^2}\,r^{2n}\,,\;\text{(Converges\;for\;small}\;r)\,.
\end{equation}
where $(a\rightarrow b)$ substitutes $a$ for $b$ in the first four terms of~\eqref{A11}.

In terms of the hypergeometric function ${}_2F_1$~\cite{Wol1}, Selvaggi \etal formula reads
\begin{equation*}
\hspace{-21mm} G(r,u,n,a,b) = b^{2n+2\,}{}_2F_1\bigg(n+1,2n+(1/2);n+2;-b^2/(u^2+r^2)\bigg) - (a\rightarrow b)\,,
\end{equation*}
\begin{equation}\label{S1}
\hspace{-21mm} V_{\text{Sel}}(r,z) = \frac{1}{2}\,\sum_{n=0}^\infty \frac{(4n-1)!!}{(n+1)(n!)^2}\,\bigg(\frac{r}{2}\bigg)^{2n}\int_{z-L}^{z+L}
\frac{G(r,u,n,a,b)}{(u^2+r^2)^{2n+(1/2)}}\,\text{d}u\,.
\end{equation}

We consider the case where $a=1/2$, $b=1$ \& $L=1$, in SI units. We restrict ourselves to the central plane $z=0$ and compare the three formulae~\eqref{L1}, \eqref{Af} \& \eqref{S1} in their respective regions of convergence. We choose three points: $P_1(0.2,0)$ near the $z$-axis, $P_2(0.4,0)$ near the inner surface of the cylinder and $P_3(2,0)$ outside the cylinder. If $V(P,n+1)$ denotes the potential at the point $P$ with $n+1$ being the number of terms used (here, $n$ represents the index of summation which has been given the same symbol in all formulae~\eqref{L1}, \eqref{Af} \& \eqref{S1}), our results follow where the exact decimal places are underlined
\begin{eqnarray}
\hspace{-15mm}\label{r1}& V_{\text{Sel}}(P_1,4) \simeq 0.\underline{820}253\, ; V_{\text{Sel}}(P_1,7) \simeq 0.\underline{82038}2\, ; V_{\text{Sch}}(P_1,3) \simeq 0.\underline{820385}\,,& \\
\hspace{-15mm}\label{r2}& V_{\text{Sel}}(P_2,5) \simeq 0.\underline{8}25570\, ;  V_{\text{Sel}}(P_2,7) \simeq 0.\underline{8}28583\, ; V_{\text{Sch}}(P_2,5) \simeq 0.\underline{831536}\,,& \\
\hspace{-15mm}\label{r3}& V_{\text{Sel}}(P_3,7) \simeq 0.\underline{371}846\, ;  V_{\text{Loc}}(P_3,7) \simeq 0.\underline{37195}8\,.&
\end{eqnarray}
$V_{\text{Sch}}(P_3)$ blows up and so do $V_{\text{Loc}}(P_1)$ \& $V_{\text{Loc}}(P_2)$. The number of exact decimal places in~\eqref{r1} to~\eqref{r3} has been identified upon comparing $V_{\text{Sch}}(P_1,3)$, $V_{\text{Sch}}(P_2,5)$ \& $V_{\text{Loc}}(P_3,7)$ to more accurate results, which have been obtained at the same points using large values of $n+1$ (see last paragraph). The relative errors at $P$ are $E_{\text{Sel-Sch}}(P_1,4)\simeq 1.61\times 10^{-4}$, $E_{\text{Sel-Sch}}(P_2,7)\simeq 3.55\times 10^{-3}$ \& $E_{\text{Sel-Loc}}(P_3,7)\simeq 3.01\times 10^{-4}$ with $E_{\text{Sel-Sch}}(P,n+1)=|V_{\text{Sel}}(P,n+1)-V_{\text{Sch}}(P,n+1)|/V_{\text{Sch}}(P,n+1)$, etc.

Near the plane $z=0$, the expansions~\eqref{L1} \& \eqref{Af} perform better than~\eqref{S1} (Eqs~\eqref{r1} to~\eqref{r3}) and are easily handled with reasonable evaluation times. The evaluation of each term in~\eqref{S1} involves a non-trivial integration which is time-consuming. In contrast, \eqref{L1} \& \eqref{Af} are evaluated via the use of Legendre polynomials, which are built-in functions in most computer algebra systems, or by differentiation. For instance, the evaluation of $V_{\text{Sel}}(P_2,7)$, by numerical integration, lasted more than 29000 times that of $V_{\text{Sch}}(P_2,7)$ (if we re-derive~\eqref{A11} using computer-algebra, this ratio reduces to 370) and the evaluation of $V_{\text{Sel}}(P_3,7)$ lasted more than 46000 times that of $V_{\text{Loc}}(P_3,7)$.

In the plane $z=0$ each term of~\eqref{S1} ($n\geq 1$ up to at least 5) reaches a maximum value for $a<r_0<b$ and decays slowly near and off the axis resulting in a slow convergence of~\eqref{S1} around $r_0$. Off the plane $z=0$, the convergence of~\eqref{S1} improves around the axis due to an off-axis shift of $r_0$ and a drop in the maximum resulting in a rapid decay of its terms ($n\geq 1$) there. At the point $P_4(0.4,2)$, we found $V_{\text{Sel}}(P_4,7)=0.\underline{36482771}$, $V_{\text{Sch}}(P_4,7)=0.\underline{36482771}$ \& $V_{\text{Loc}}(P_4,7)=0.\underline{36482}807$. The evaluation of $V_{\text{Sel}}(P_4,7)$ lasted 8500 times that of $V_{\text{Sch}}(P_4,7)$ and 52000 times that of $V_{\text{Loc}}(P_4,7)$.

The flexibility in using~\eqref{L1} or \eqref{Af} allowed as to evaluate the GP using up to forty-one terms and to check over the stability of the results shown in~\eqref{r1} to~\eqref{r3} for $V_{\text{Sch}}$ \& $V_{\text{Loc}}$. Moreover, the GP at each of the points $P_1$ to $P_4$ has been reobtained by numerically integrating the volume integral of the potential element $\text{d}v'/(2\pi |\mathbf{r}-\mathbf{r'}|)$.

\end{document}